\documentclass{aa}  
\usepackage{graphicx}
\usepackage{txfonts}
\usepackage{soul}
\usepackage{hyperref}
\hypersetup{
    colorlinks=true,
    citecolor=blue,
    urlcolor=blue
    }
    
\usepackage{multirow}
\usepackage{threeparttable, tablefootnote}
\usepackage{placeins}

\graphicspath{{images/}}

\newcommand{\Msun}{\rm M_{\odot}}
\newcommand{\MsunYr}{\rm M_{\odot}~yr^{-1}}

\newcommand{\mum}{\mu \rm m}

\begin{document}

   %\title{Modeling submillimeter galaxies in cosmological simulations: Evolution in large-scale environment}
   %\title{Evolution of submillimeter galaxies in different cosmic environments}
   \title{Evolution of submillimeter galaxies across cosmic-web environments}

   %\subtitle{I. Environment}
   \titlerunning{Cosmic environment of submillimeter galaxies} 
   \authorrunning{Kumar et al. 2025}

   \author{Ankit Kumar\inst{1}\fnmsep\thanks{ankit4physics@gmail.com}
          \and
          M. Celeste Artale\inst{1}\fnmsep\thanks{maria.artale@unab.cl}
          \and
          Antonio D. Montero-Dorta\inst{2}
          \and
          Lucia Guaita\inst{1,3}
          \and
          Joop Schaye\inst{4}
          \and
          Kyoung-Soo Lee\inst{5}
          \and
          Alexandra Pope\inst{6}
          \and
          Facundo Rodriguez\inst{7,8}
          \and
          Eric Gawiser\inst{9}
          \and
          Ho Seong Hwang\inst{10,11}
          \and
          Paulina Troncoso Iribarren\inst{12}
          \and
          Jaehyun Lee\inst{13}
          \and
          Seong-Kook Lee\inst{11}
          \and
          Changbom Park\inst{14}
          \and
          Yujin Yang\inst{13}
          %+++ Daniela ?
          }

   \institute{Universidad Andres Bello, Facultad de Ciencias Exactas, Departamento de Fisica y Astronomia, Instituto de Astrofisica, Fernandez Concha 700, Las Condes, Santiago RM, Chile
              %\email{ankit4physics@gmail.com}
         \and
         Departamento de Física, Universidad Técnica Federico Santa María, Avenida Vicuña Mackenna 3939, San Joaquín, Santiago, Chile
            %\email{email2}
        \and
        Millennium Nucleus for Galaxies (MINGAL)
        \and
         Leiden Observatory, Leiden University, PO Box 9513, 2300 RA Leiden, the Netherlands
         \and
         Department of Physics and Astronomy, Purdue University, 525 Northwestern Avenue, West Lafayette, IN 47907, USA
         %\email{email3}
            %\thanks{Just to show the usage of the elements in the affiliation field}
         \and
         Department of Astronomy, University of Massachusetts, Amherst, MA 01003, USA
         \and
         CONICET. Instituto de Astronomía Teórica y Experimental (IATE). Laprida 854, Córdoba X5000BGR, Argentina
         \and
         Instituto de Astronomía Teórica y Experimental (IATE), CONICET-UNC, Laprida 854, X500BGR, Córdoba, Argentina
         \and
         Physics and Astronomy Department, Rutgers, The State University, Piscataway, NJ 08854, USA
         \and
         Department of Physics and Astronomy, Seoul National University, 1 Gwanak-ro, Gwanak-gu, Seoul 08826, Republic of Korea
         \and
         SNU Astronomy Research Center, Seoul National University, 1 Gwanak-ro, Gwanak-gu, Seoul 08826, Republic of Korea
         %\and
         %Australian Astronomical Optics - Macquarie University, 105 Delhi Road, North Ryde, NSW 2113, Australia
         \and
         Facultad de Ingeniería y Arquitectura, Universidad Central de Chile, Chile
         \and
         Korea Astronomy and Space Science Institute, 776 Daedeokdae-ro, Yuseong-gu, Daejeon 34055, Republic of Korea
         \and
         Korea Institute for Advanced Study, 85 Hoegi-ro, Dongdaemun-gu, Seoul 02455, Republic of Korea
             }

   % \date{Received September 15, 1996; accepted March 16, 1997}

% \abstract{}{}{}{}{} 
% 5 {} token are mandatory
 
  \abstract
  % context heading (optional)
  % {} leave it empty if necessary  
   {
   Submillimeter galaxies (SMGs) are
   among the most intriguing and luminous objects in the high-redshift universe. These highly dusty, star-forming systems provide valuable insights into galaxy formation and evolution and are likely influenced by their cosmic environment. Observationally, however, their rarity makes environmental trends difficult to establish. %In this context, the limited number of observed SMGs %and the difficulty of reproducing their properties in large cosmological hydrodynamical simulations 
   %make it challenging to investigate their cosmic environment.
   }
  % aims heading (mandatory)
   {
    We use the large-volume cosmological hydrodynamical simulation FLAMINGO, which simultaneously reproduces the redshift distribution and number counts of SMGs without invoking a top-heavy initial mass function to investigate their evolution in different cosmic environments.
   }
  % methods heading (mandatory)
   {
   We use the DisPerSE cosmic web finder to identify filamentary structures at redshifts $z=4$, 3, 2, 1.5, and 1 in the highest-resolution FLAMINGO simulation using galaxies with M$_{*} \geq 10^{9}~\Msun$ as tracers of large-scale structure. %Additionally, we test our findings using a fixed number density approach.
   We define five cosmic environments (inner cluster-halo, outer cluster-halo, inner filament, outer filament, and void/wall) at each redshift considering mass evolution of cluster-halos and density evolution of filaments.
   }
  % results heading (mandatory)
   {
   For a fixed stellar-mass cut of $M_* \geq 10^{9}~\Msun$, the fraction of SMGs in the inner cluster-halo environment declines from $\sim30\%$ at $z=4$ to $\sim3\%$ by $z=1$, and similar trends are observed in other environments. 
   The abundance of SMGs within a cluster-halo increases with halo mass, mirroring the increase in the total galaxy population. Consequently, the ratio of SMG halo occupation to that of all galaxies is largely insensitive to halo mass, but varies with redshift. 
   In contrast, the ratio of the halo occupation of non-SMGs to that of all galaxies declines with halo mass and shows little redshift evolution due to suppressed star formation in massive halos. 
   We further show that SMGs tend to contribute less to the massive end of the stellar mass function in low-density environments, with central and satellite SMGs forming two distinct populations in inner cluster-halos. 
  SMGs occupy the metal-rich side of the star-forming gas metallicity distribution, but rarely attain the very highest metallicities because ongoing enrichment is limited by gas depletion.
  Interestingly, the brightest SMGs (S$_{850} > 10$ mJy) are found exclusively in inner cluster-halos, highlighting a strong connection between SMG luminosity and environmental density. Our results show that SMGs dominate star formation in dense environments, contributing up to $80\%$ of the SFR in inner cluster-halos at $z=4$, but less than $50\%$ in low-density regions.
  Taken together, these results demonstrate that SMGs dominate the build-up of stellar mass and metals in dense environments, while their relative importance declines towards lower redshift and lower-density regions.
   }
  % conclusions heading (optional), leave it empty if necessary 
   {}

   \keywords{Galaxies: formation -- Galaxies: evolution -- Submillimeter: galaxies -- Infrared: galaxies -- Galaxies: high-redshift}

   \maketitle
%
%-------------------------------------------------------------------

\section{Introduction}
\label{sec:introduction}
Submillimeter galaxies (SMGs) are a population of high-redshift, heavily dust-obscured galaxies undergoing intense star formation, with star formation rates (SFRs) often exceeding $100$--$1000~\MsunYr$ \citep[e.g.,][]{Barger.etal.1998, Chapman.etal.2005, Barger.etal.2012, Hezaveh.Marrone.etal.2013, Swinbank.etal.2014}. These galaxies are typically identified through their thermal emission in the submillimeter wavelength range, where dust heated by young and massive stars re-emits absorbed ultraviolet and optical light. SMGs are largely invisible in optical surveys due to their dust content, but appear as some of the most luminous objects at submillimeter and far-infrared wavelengths. Their redshift distribution peaks around $z \simeq 2$--$3$, aligning with the peak epoch of cosmic star formation activity \citep{Chapman.etal.2005, Swinbank.etal.2014, Danielson.etal.2017, Dudzeviciute.etal.2020}. 

One of the most compelling observational advantages of SMGs is the effect of the negative $K$-correction, which causes their observed flux density to remain nearly constant across a broad redshift range ($z \simeq 0.5$--$10$) in the submillimeter regime \citep{Blain.1997}. This property makes SMGs particularly well-suited for detecting distant galaxies that would otherwise be missed in optical surveys. This makes SMGs a critical population for studying the buildup of stellar mass and the evolution of massive galaxies in the early Universe \citep{Smail.etal.1997, Barger.etal.1998, Hughes.etal.1998, Blain.etal.2002, Negrello.etal.2010}.

SMGs exhibit molecular gas masses of $M_{\rm g} \gtrsim~10^{10}~\Msun$, as inferred from CO line observations \citep{Frayer.etal.1999, Tacconi.etal.2008, Bothwell.etal.2013}, median stellar masses of $\approx 10^{11}\Msun$ \citep{Hainline.etal.2011, Michalowski.eta.2012, da_Cunha.etal.2015, Michalowski.etal.2017, Miettinen.etal.2017}, and halo masses of $\approx 10^{13}\Msun$ derived from spatial clustering \citep{Chen.etal.2016, Lim.etal.2020}. Their dust masses, often $\approx 10^8\Msun$, imply efficient dust production and enrichment within short cosmic timescales. Submillimeter interferometry has revealed compact starburst cores with sizes of $\approx 1$–$2$ kpc \citep{Simpson.etal.2015, Hodge.etal.2016}, suggesting that their extreme SFRs are fueled by concentrated gas reservoirs. Many SMGs exhibit disturbed morphologies and kinematics \citep{Conselice.etal.2003, Pope.etal.2005}, interpreted as evidence for major mergers \citep{Frayer.etal.1999, Ivison.etal.2000, Engel.etal.2010, Alaghband-Zadeh.etal.2012}, although some systems show signatures of rotationally supported disks \citep{Greve.etal.2005, Tacconi.etal.2008, Hodge.etal.2012, Chen.etal.2017}.

Theoretical models have explored various formation channels to explain the origin of SMGs. Early semi-analytical models (SAMs) required top-heavy initial mass functions (IMFs) to simultaneously reproduce the observed redshift distribution and number counts of SMGs \citep{Baugh.etal.2005, Lacey.etal.2016}. More recent SAMs, however, have achieved reasonable agreement with observations without invoking a top-heavy IMF \citep{Safarzadeh.etal.2017, Lagos.etal.2020}. Parametric models based on radiative transfer post-processing of high-resolution simulations \citep{Hayward.etal.2011, Hayward.etal.2013, Lovell.etal.2021, Cochrane.etal.2023} have also been employed to estimate the submillimeter fluxes of model galaxies \citep{Hayward.etal.2021, Araya-Araya.etal.2024}. Nevertheless, reproducing both the observed number counts and redshift distributions of SMGs simultaneously remains a significant challenge. Full cosmological hydrodynamical simulations have historically struggled to capture the SMG population accurately, although recent simulations show marked improvements \citep{Shimizu.etal.2012, McAlpine.etal.2019, Lovell.etal.2021}. Notably, \citet{Kumar.etal.2025} successfully reproduced both the redshift distribution and number counts of SMGs within the FLAMINGO simulation \citep{Schaye.etal.2023, Kugel.etal.2023}.

SMGs are believed to play a central role in the formation of massive elliptical galaxies. Their high SFRs, stellar masses, and short gas depletion timescales (few hundreds Myr) suggest that they represent a short-lived but important phase in the rapid assembly of massive galaxies \citep{Fu.etal.2013, Simpson.etal.2014, Toft.etal.2014, Wilkinson.etal.2017, Gomez-Guijarro.2018, Valentino.etal.2020}. Observations of post-starburst galaxies at $z \approx 1$--$2$ support an evolutionary connection between SMGs and compact quiescent galaxies \citep{Barro.etal.2014, Williams.etal.2021}. Feedback from active galactic nuclei (AGN) or supernovae may quench star formation and terminate the SMG phase, facilitating the transition from dusty starbursts to red and dead ellipticals \citep{Narayanan.etal.2010, Whitaker.etal.2021}.

The environments of SMGs offer valuable insights into their formation and evolution. Large-scale structure analyses have explored whether SMGs preferentially reside in dense regions such as proto-clusters, filaments, or massive dark matter halos. Studies of SMG overdensities have revealed associations with known galaxy overdensities and proto-clusters at $z \gtrsim 2$ \citep{Ivison.etal.2013, Casey.etal.2015, Umehata.etal.2015, Miller.etal.2018, Harikane.etal.2019, Lagache.etal.2026}, although not all SMGs are found in such environments \citep{Cornish.etal.2024}. Estimates of the fraction of SMGs residing in over-dense regions vary widely, ranging from $\approx 30\%$ to $60\%$ \citep{Smolvcic.etal.2017, Alvarez-Crespo.etal.2021}, suggesting a diversity of formation pathways influenced by both local and large-scale environments. Cosmological simulations suggest that a significant fraction of SMGs lie along filamentary structures that channel cold gas into massive halos, thereby fueling sustained star formation \citep{Dekel.etal.2009, Narayanan.etal.2015}. The influence of cosmic web structures—such as voids, sheets, filaments, and nodes—on SMG formation and evolution remains an active area of research. In recent years, cosmic web classification techniques have advanced considerably, enabling detailed analyses of galaxy environments. Applying these tools to SMGs provides a systematic approach to assess whether they preferentially occupy gas-rich filaments, merger-prone nodes, or dynamically cold regions like sheets. Understanding the spatial correlation between SMGs and the cosmic web is crucial for disentangling the physical processes that trigger their intense starburst activity.

In addition to direct environmental associations, clustering analyses suggest that SMGs at $z \geq 2$ reside in halos with masses of $M_{\rm halo} \sim 10^{12.5}$–$10^{13.5}~M_\odot$ \citep{Hickox.etal.2012, Chen.etal.2016, Wilkinson.etal.2017, Amvrosiadis.etal.2019}, placing them among the most massive galaxies of their epoch and identifying them as likely progenitors of today’s brightest cluster galaxies.  These studies also report relatively large correlation lengths for SMGs, typically $r_{0} \approx 6$--$10~ h^{-1}$ Mpc, indicating strong spatial clustering. Such large $r_{0}$ values suggest that SMGs are preferentially located in dense environments within the cosmic web, such as the intersections of filaments and %or the outskirts of 
proto-clusters, further supporting their association with massive dark matter halos and their role in tracing early large-scale structure \citep{Blain.etal.2004, Chen.etal.2016, Wilkinson.etal.2017}.

In light of large-scale surveys of SMGs--including SCUBA \citep{Smail.etal.1997, Hughes.etal.1998}, SCUBA-2 \citep{Chapman.etal.2005, Coppin.etal.2006, Geach.etal.2017}, LABOCA \citep{Siringo.etal.2009, Weiss.etal.2009}, AzTEC \citep{Scott.etal.2008, Hatsukade.etal.2011}, Herschel \citep{Eales.etal.2010, Oliver.etal.2012}, ALMA \citep{Hodge.etal.2013, Karim.etal.2013, Stach.etal.2018}, and TolTEC \citep{Wilson.etal.2020}--we aim to investigate the connection between SMGs and different cosmic environments using the large-volume cosmological simulation FLAMINGO, which has been shown to reproduce observed properties of SMGs \citep{Kumar.etal.2025}. Furthermore, we compare SMGs with a submillimeter-faint star-forming galaxy sample, intended to broadly resemble UV-selected populations such as Lyman-$\alpha$ emitters (LAEs) and Lyman-break galaxies (LBGs), commonly used as tracers of large-scale structure at high redshift \citep{Ito.etal.2020, Lee.etal.2024, Andrews.etal.2024, Ramakrishnan.etal.2024}.

This paper is organized as follows. Section~\ref{sec:simulation} provides a brief overview of the cosmological hydrodynamical simulation, FLAMINGO, used in this work. In Section~\ref{sec:smg_modeling}, we describe the modeling of the SMG population within cosmological simulations. Section~\ref{sec:disperse} discusses the identification of the cosmic web using DisPerSE. Section~\ref{sec:cosmic_env} outlines our definitions of different cosmic environments. Section~\ref{sec:results} presents our results, including the distribution of SMGs across various cosmic environments, their halo occupation, their contribution to the cosmic star formation rate density in different environments, and their physical properties. Finally, Section~\ref{sec:conclusions} summarizes the main findings of this work.

%--------------------------------------------------------------------
\section{The FLAMINGO hydrodynamical simulation}
\label{sec:simulation}
In this work, we use the FLAMINGO\footnote{\url{https://flamingo.strw.leidenuniv.nl/}} (Full-hydro Large-scale structure simulations with All-sky Mapping for the Interpretation of Next Generation Observations) simulation of the Virgo Consortium\footnote{\url{https://virgo.dur.ac.uk/}}. The FLAMINGO is a suite of large-volume hydrodynamical cosmological simulations performed with the \textsc{Swift} code \citep{Schaller.etal.2024} starting from redshift $z=31$ and run down to $z=0$. It uses several new and improved subgrid prescriptions. These prescriptions include element-by-element radiative cooling and heating \citep{Ploeckinger.Schaye.2020}, star formation \citep{Schaye.Dalla_Vecchia.2008}, stellar mass-loss from massive stars, asymptotic giant branch stars, core-collapse supernovae and type Ia supernovae \citep{Wiersma.etal.2009, Schaye.etal.2015}, kinetic stellar feedback from massive stars and supernovae \citep{Chaikin.etal.2023}, super massive black hole seeding and evolution \cite{Springel.2005, Booth.Schaye.2009, Bahe.etal.2022}, AGN feedback \citep{Booth.Schaye.2009}, and various other improvements. It uses the \cite{Chabrier.etal.2003} IMF to model stellar evolution. For more details on the simulations, we refer the reader to \cite{Schaye.etal.2023}. The subgrid physics in the FLAMINGO simulations is calibrated to match only two observables: the gas fraction in low-redshift galaxy clusters and the galaxy stellar mass function at $z=0$. In contrast to traditional trial and error methods, FLAMINGO performed these calibrations using machine learning \citep{Kugel.etal.2023}.

FLAMINGO presents four fiducial hydrodynamical runs: three in a (1 cGpc)$^{3}$ box and one in a (2.8 cGpc)$^{3}$ box. 
%They are identified as L1\_m8, L1\_m9, L1\_m10, and L2p8\_m9 according to the box side length and the rounded baryonic particle mass on a logarithmic scale with base 10. The high-resolution box (L1\_m8) and large box (L2p8\_m9) runs are the flagship simulations of the FLAMINGO. Additionally, there are eight variations for galaxy formation prescriptions and four variation for cosmologies, all in (1 cGpc)$^{3}$ box. The data is stored in 79 snapshots from redshifts $z=15$ to $z=0$. 
The halos are identified using the Friends-of-Friends \citep[\textsc{FoF}:][]{Press.Davis.1982, Davis.etal.1985} algorithm on all particles (except neutrinos). FLAMINGO uses the \textsc{hbt-herons} \citep[Hierarchical Bound Tracing - Hydro-Enabled Retrieval of Objects in Numerical Simulations:][]{Han.etal.2018, Forouhar_Moreno.etal.2025} algorithm for \textsc{FoF} implementation and for the gravitationally bound subhalos (galaxies) identification. The halos and subhalos are processed using the Spherical Overdensity and Aperture Processor \citep[SOAP,][]{McGibbon.etal.2025} to measure various properties in 3D or projected apertures.

For this study, we use the highest-resolution FLAMINGO run, namely L1\_m8. The initial masses of baryonic and dark matter particles in L1\_m8 are, respectively $1.34 \times 10^{8}~\Msun$ and $7.06 \times 10^{8}~\Msun$. It has a maximum gravitational softening length of 2.85 physical kpc (pkpc), while the comoving softening is held constant for $z>2.91$. The number of stellar and dark matter particles is the same, and each is $1.8^{3}$ times the number of neutrinos. The full box consists of $2 \times 3600^{3} \text{(dark matter and gas)} +2000^{3} \text{(neutrino)}$ resolution elements. We consider only those galaxies which have at least ten stellar particles, i.e., $\rm M_{*} > 10^{9}~\Msun$ down to which the observed stellar mass function is reproduced \citep{Schaye.etal.2023}.

\section{Modeling submillimeter flux densities}
\label{sec:smg_modeling}
Modeling submillimeter flux densities for galaxies in hydrodynamical simulations is a challenging and computationally expensive task, especially when it comes to large-volume simulations where the ISM is not sufficiently resolved. In such large simulations with unresolved ISM, semi-empirical or parametric relations are required \citep[see e.g.,][]{Shimizu.etal.2012,Dave.etal.2010,Hayward.etal.2021}. Recently, in \cite{Kumar.etal.2025}, we tested various parametric relations to model the submillimeter fluxes in different cosmological simulations. We found that by applying the \cite{Hayward.etal.2013} parametric relation to FLAMINGO we can reasonably reproduce the observed submillimeter flux densities.
 %\citep[see,][]{Kumar.etal.2025}. 

The parametric relation reported in \citet{Hayward.etal.2013} for modeling the SCUBA\_2 850 $\mum$ flux densities is expressed as,
\begin{equation}
    S_{850}~[\mathrm{mJy}] = {0.81} \left(\frac{\mathrm{SFR}}{\mathrm{100~\MsunYr}}\right)^{0.43} \left(\frac{M_{\mathrm{dust}}}{\mathrm{10^{8}~\Msun}}\right)^{0.54},
    \label{eqn:s850_relation}
\end{equation}
where SFR is the instantaneous star formation rate, and $M_{\rm dust}$ is the dust mass. For the submillimeter flux density calculation, the instantaneous SFR is readily available for hydrodynamical simulations and the dust mass can be obtained from the metal mass assuming a constant dust-to-metal (DTM) ratio.

We first compute the SFR and $M_{\rm dust}$ properties of each subhalo (i.e., galaxy). For the SFR calculation, we use a spherical aperture of 50 physical kpc (pkpc) centered on the subhalo center, and sum the instantaneous SFRs of each star-forming gas particle within the aperture. For dust mass, we assume a fixed dust-to-metal (DTM) mass ratio of 0.4 following observational estimates from \cite{Dwek.1998} and \cite{James.etal.2002}. Further, we consider only cold star-forming gas particles for the metal mass calculation because thermal sputtering and collisions in the hot gas destroy dust grains \citep{Draine.Salpeter.1979, McKinnon.etal.2016, Popping.etal.2017}. By virtue of the star formation criterion, all the gas particles that are forming stars are cold and dense for the metal mass calculation. The sum of the metal mass in all the cold star-forming gas particles returns the total metal mass of the galaxy, which is converted into a dust mass using DTM $=0.4$. The exact DTM ratio does not affect the measured flux densities significantly. Using DTM $=0.5$ would increase the flux densities by only $13\%$.

In this work, we refer to all the galaxies with S$_{850} > 1$ mJy as SMGs. Our choice of flux density selects relatively bright SMGs that are easily detectable in recent and upcoming large submillimeter surveys. Additionally, we also select a population of submillimeter-faint star-forming galaxies to put in the context of other potential tracers of large-scale structure, e.g., bright LAEs, LBGs. 
This sample is not constructed to reproduce LAE/LBG selection functions, but it is meant to capture a population that is qualitatively similar (i.e., actively star-forming and relatively dust-poor, \citealt{Eric.etal.2007, Coppin.etal.2015, Firestone.etal.2025}) to UV-selected galaxies such as LAEs and LBGs.
%Since LAEs and LBGs are star-forming galaxies and show little dust obscuration \citep{Eric.etal.2007, Coppin.etal.2015, Firestone.etal.2025}, we select this population as having SFR > 2 $\MsunYr$ and S$_{850} < 0.5$ mJy. We refer to these galaxies as non-SMGs.

\section{Cosmic web identification for SMGs}
In this section, we outline the approach used to detect cosmic web structures and their association with SMGs. Our analysis relies on identifying filamentary structures and massive halos across different redshifts. To trace the filaments, we employ the DisPerSE filament-finding algorithm. Here we briefly outline how DisPerSE is implemented in this work, and we then describe the different cosmic web components considered in our analysis.

\subsection{The DisPerSE filament finder on the FLAMINGO simulation}\label{sec:disperse}
%\section{Cosmic web identification using DisPerSE}}

%Simulations benefit from detailed information on the distribution of gas, stars, and dark matter, to use them as the cosmic web tracers. Gas in the vicinity of halos is subject to strong stellar and AGN feedback, while dark matter observations are not yet feasible. Thus, for the practical consistency with several observational studies, we use stellar component of galaxies as the cosmic web tracer. We consider galaxies with M$_{*} \ge 10^{9}~\Msun$, ensuring observational limits and sufficient number of stellar particles in a simulated galaxy.

We use the Discrete Persistent Structures Extractor \citep[DisPerSE,][]{Sousbie..2011, Sousbie.etal.2011} code
%\footnote{\url{https://www2.iap.fr/users/sousbie/web/html/indexd41d.html?}}
to identify the filamentary structures in the FLAMINGO simulation. 
Several other algorithms exist for identifying the filamentary network \citep[see e.g.,][]{Novikov2006, Aragon2007, Cautun2013, Burchett2020}. Compared to other approaches, DisPerSE operates directly on point distributions offers a topology-based framework that is less sensitive to arbitrary smoothing scales and is particularly well suited for consistently tracing filaments across redshift.

DisPerSE has been applied to both observational data \citep[see e.g.,][]{Ramakrishnan.etal.2024, Rodriguez2026} and cosmological simulations \citep[see e.g.,][]{Lee.etal.2021, Galarraga2022, Zakharova.etal.2023, Perez2024, MonteroDorta2024, Bahe2025}. In simulations, it can further exploit the detailed information of gas, stars, and dark matter, which serve as tracers of the cosmic web and, in particular, allow for the robust identification of filamentary structures \citep[see e.g.,][]{Galarraga2022,Im2024,Bahe2025,Navdha2025}.
Some recent works have further investigated the consistency between using galaxies and dark matter as tracers of the cosmic web \citep{Bahe2025}.
%Gas in the vicinity of halos is subject to strong stellar and AGN feedback, while dark matter observations are not yet feasible. Thus, for the practical consistency with several observational studies, we use stellar component of galaxies as the cosmic web tracer. We consider galaxies with M$_{*} \ge 10^{9}~\Msun$, ensuring observational limits and sufficient number of stellar particles in a simulated galaxy.
Since our goal is to mimic observational trends, we run DisPerSE using galaxies with M$_{*} \ge 10^{9}~\Msun$ as tracers of the filamentary structures for all the redshifts studied. %We further test our results by tracing filaments with a constant comoving galaxy number density, following \citet{Galarraga-Espinosa.etal2024}. 

Although DisPerSE can handle periodic boundary conditions of the simulation box, it struggles with the sparse discreteness, like in our case, where we use galaxies as tracers. To mitigate this, we extend the box by 20\% in each direction by repeating it along all axes. Starting from the discrete galaxy distribution, we compute the density field at the location of tracers using the \texttt{delaunay\_3D} program of DisPerSE, which uses the delaunay tessellation field estimator \citep[DTFE,][]{Schaap.van-de-Weygaert.2000}. We use the \texttt{netconv} program of DisPerSE for smoothing the density field once to reduce shot noise (also see \citealt{Galarraga-Espinosa.etal2024}). It smooths density field by replacing each vertex’s value with the average of its connected neighbors. Then the \texttt{mse} program of the DisPerSE uses this density field to extract the critical points (voids, saddles, nodes) and filament segments using discrete Morse theory. We execute the \texttt{mse} program with 2 sigma persistence. Finally, we run the \texttt{skelconv} program of DisPerSE to convert the binary format output of the \texttt{mse} program to the human-readable format for critical points and segments
\footnote{DisPerSE commands and flags used in this work:\\
\texttt{\$ delaunay\_3D}\\
\texttt{\$ netconv -smoothData field\_value 1}\\
\texttt{\$ mse -nsig 2 -forceLoops -manifolds -upSkl}\\
\texttt{\$ skelconv -breakdown -smooth 1 -to crits\_ascii}\\
\texttt{\$ skelconv -breakdown -smooth 1 -to segs\_ascii}
}.

%before: \subsection{Defining cosmic environment of the SMGs}
\subsection{Definition of the cosmic environments}
\label{sec:cosmic_env}

The evolution and formation of cosmic web structures occurs throughout cosmic time. For instance, employing a constant halo mass threshold from $z=0$ clusters to define cluster-equivalent environment (hereafter referred to as cluster-halo environments)\footnote{We use the term `cluster-halo environment' to avoid confusing it with `cluster environment' and `halo environment.' At high redshift, halos are not relaxed enough to be considered clusters.} is ineffective at high redshifts, as halos have yet to accumulate sufficient mass. The most massive halo in the FLAMINGO simulation at $z=3$ has M$_{200}=6.8~\times~10^{13}~\Msun$, whereas at $z=0$ clusters are typically defined to have M$_{200}~\ge~10^{14}~\Msun$. Thus, we should consider an evolving picture of the cosmic web when associating SMGs with different environments at different redshifts. 

%To define the cluster-halo environment at different redshifts, we begin with redshift zero and consider $z=0$ halos with M$_{200} \ge 10^{14}~\Msun$ as cluster-halos. 
We characterize the cluster-halo environment by identifying, at redshift zero, all halos with M$_{200} \ge 10^{14}~\Msun$ as cluster-halos.
There are 9736 cluster-halos in the FLAMINGO simulation at $z=0$, which corresponds to a cluster-halo number density of $9.736~\times~10^{-6}$ cMpc$^{-3}$. We use this comoving cluster-halo number density to infer the minimum halo mass of the cluster-halos at any other redshift, assuming it remains constant during the evolution. Thus, at any given redshift, the minimum halo mass of the cluster-halos corresponds to the mass of the least massive halo required to achieve a cluster-halo number density of $9.736~\times~10^{-6}$ cMpc$^{-3}$. This translates to minimum halo masses of M$_{200} = 4.3~\times~10^{13}$, $2.6~\times~10^{13}$, $1.6~\times~10^{13}$, $6.5~\times~10^{12}$, and $3.0~\times~10^{12}$ $\Msun$ at $z=1$, 1.5, 2, 3, and 4, respectively. Though massive halos at any given redshift do not necessarily evolve into $z=0$ cluster-halos, our assumption provides a crude approximation for incorporating mass build-up while defining the cluster-halo environment across the redshift range. We define the `inner cluster-halo' members as those within an R$_{200}$ radial distance of a cluster-halo, and the `outer cluster-halo' members as those between R$_{200}$ and 3R$_{200}$ radial distances. 

\begin{figure*}[!ht]
    \centering
    \includegraphics[width=\textwidth]{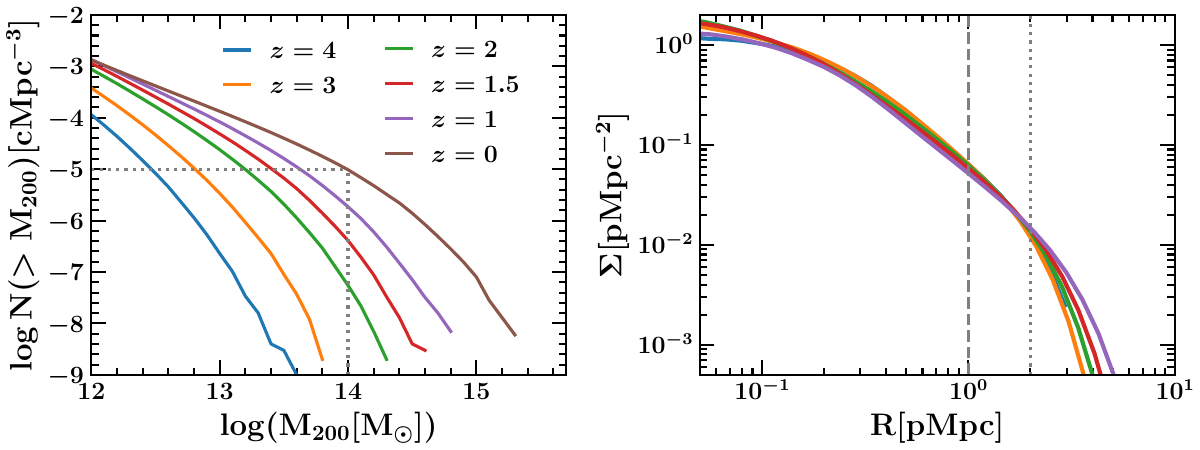}
    \caption{\textit{Left:} panel shows the cumulative halo mass functions at redshifts $z=4$, 3, 2, 1.5, 1, and 0, respectively, with the blue, orange, green, red, purple, and brown color curves. The vertical gray color dotted line indicates the minimum halo mass at $z=0$ used to define cluster-halo, whereas horizontal line indicate corresponding halo number density at $z=0$. The intersections of horizontal gray color line with cumulative halo mass functions indicate the lower mass limit of halo ($\rm M_{Thresh} (z)$) used to define cluster-halo at corresponding redshift. \textit{Right:} panel shows the stacked radial galaxy number density for the filaments collapsed along their spines. The vertical dashed and dotted gray color lines indicate the radii used to define the extent of inner ($R = 1$ pMpc) and outer ($R = 2$ pMpc) filament regions, respectively.}
    \label{fig:env_definition}
\end{figure*}

In the left panel of Fig.~\ref{fig:env_definition}, we show the redshift evolution of the cumulative halo mass function from $z=4$ to $z=0$. The vertical gray dotted line marks the lower halo mass limit (M$_{200}=10^{14}~\Msun$) for cluster-halo definition at $z=0$. The horizontal gray dotted line indicates the cumulative number density at $z=0$ that we use corresponding to define cluster-halo environments at high redshift. There is no halo at $z \geq 3$ that satisfies the $z=0$ cluster-halo mass definition. 
A redshift-independent (fixed) mass threshold such as $M_{200}\ge 10^{14}\,M_\odot$ would yield too few (or zero) objects at high redshift, preventing robust statistics. We therefore adopt a constant comoving number density selection to define `cluster-halo' environments across redshift, which approximately tracks similarly rare overdensities over cosmic time.
%Therefore, we opt to use a constant comoving cluster-halo number density at all redshifts to define cluster-halos.

Similarly, to define the filament environment at any redshift, we calculate the stacked radial galaxy number density for the filaments collapsed along their spines. The stacked radial density $\Sigma (R)$ at any radius $R$ from the center of collapsed filaments is measured in the annular region of radial width $2\Delta R$ using the following expression.
\begin{equation}
    \Sigma(R) = \frac{1}{N_{\rm fil}}\frac{N(R)}{\pi[(R+\Delta R)^{2}-(R-\Delta R)^{2}]},
\end{equation}
where $N(R)$ is the number of galaxies with $M_{*} \ge 10^{9}~\Msun$ in the annular region at radius $R$, and $N_{\rm fil}$ is the total number of galaxies whose nearest structure is the filament spine. Since the filaments are connected to the nodes, some galaxies in the cluster-halos are closer to the filaments than to the nodes, particularly galaxies in the `outer cluster-halo' region. We removed all the cluster-halo members within a 3R$_{200}$ radial distance while computing the stacked radial galaxy number density profile of filaments to avoid any contamination from cluster-halos. 

The right panel of Fig.~\ref{fig:env_definition} shows the stacked radial galaxy number density profiles for filaments collapsed along their spines at $z=4$, 3, 2, 1.5, and 1. % with blue, orange, green, red, and purple color curves, respectively. 
The stacked radial galaxy number density profiles are similar across these redshifts.
%It is clear that the stacked radial galaxy number density profiles remain nearly the same at all the redshifts shown here.
In view of the minimal evolution of the stacked density profiles of filaments in physical space, we define `inner filament' members as those within 1 pMpc distance from a filament spine ($D_{Skeleton}$), and `outer filament' members as those between 1 and 2 pMpc distance from a filament spine. The vertical dashed and dotted gray lines in the right panel of Fig.~\ref{fig:env_definition} indicate the extent of the inner and outer filament regions.
Our choice for inner and outer filament radii is also motivated by previous work using DisPerSe-based filament reconstructions in simulations \citep[see e.g.,][]{GalarragaEspinosa2023}.

After defining cluster-halo and filament environments, we consider remaining galaxies as members of voids and walls. We collectively define these two environments as the `void/wall' environment. To summarize, we define five environments as follows:

\begin{flushleft}
\begin{tabular}{ll}
\text{Inner cluster-halo:} & $\rm M_{200} \ge M_{Thresh}(z) \, \text{ and } \, R \le R_{200}$ \\
\text{Outer cluster-halo:} & $\rm M_{200} \ge M_{Thresh}(z) \, \text{ and } \, R_{200} < R \le 3R_{200}$ \\
\text{Inner filament:} & $\rm D_{Skeleton} \le 1 \, \text{pMpc}$ \\
\text{Outer filament:} & $\rm 1 \, \text{ pMpc} < D_{Skeleton} \le 2 \, \text{ pMpc}$ \\
\text{Void/wall:} & Rest of the galaxies \\
\end{tabular}
\end{flushleft}

\section{Results}
\label{sec:results}
We present our results from the FLAMINGO simulation.
% L1\_m8, and refer to it as FLAMINGO. We discuss the redshift evolution of SMGs and their properties in different cosmic environments.
In Section~\ref{sec:fractional_evolution} we investigate the fraction of SMGs in the different cosmic environments, and the SMG halo occupancy in the inner cluster-halos. We then investigate the SMG properties in Section~\ref{sec:properties_SMG}, and finally, explore the contribution to the star formation rate density (SFRD) of the SMGs in different cosmic web environments in Section~\ref{sec:contribution_to_SFRD}.

\subsection{The evolution of the SMGs fraction }
\label{sec:fractional_evolution}

\begin{figure}
    \centering
    \includegraphics[width=\columnwidth]{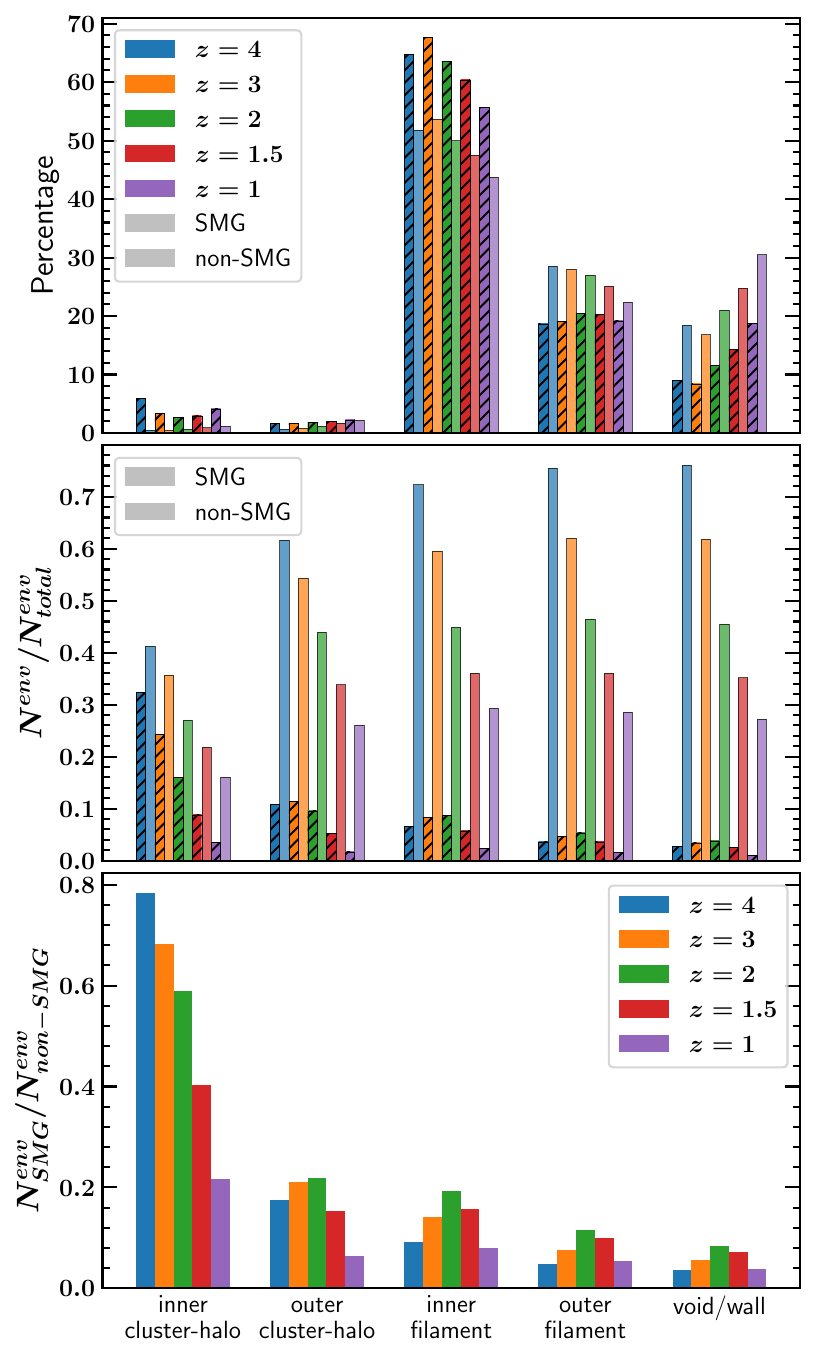}
    \caption{\textit{Top:} The percentage of SMGs (non-SMGs) in different environments using hatched (unhatched) bars. Redshifts $z=4$, 3, 2, 1.5, and 1 are represented by the blue, orange, green, red, and purple colors, respectively. %As expected, most galaxies reside in filamentary environments. 
    \textit{Middle:} Redshift evolution of the SMG and non-SMG fractions across cosmic environments, normalized by the total galaxy population in each environment. %At a given redshift, SMGs (non-SMGs) tend to decrease (increase) when moving towards low-density environment. 
    \textit{Bottom:} Redshift evolution of the SMG-to-non-SMG ratio in different cosmic environments. In all three panels, SMGs and non-SMGs are modeled for galaxies with $\rm M_{*} \ge 10^{9}~\Msun$.}
    \label{fig:fractional_evolution}
\end{figure}

%Filamentary environments are known to host the bulk of galaxies in the universe \cite{Cautun.etal.2014, Ganeshaiah_Veena.etal.2019, Pandey.Nandi.2025}. Building on this, 
In the top panel of Fig.~\ref{fig:fractional_evolution}, we show the percentages of SMGs (hatched bars) across environments at redshifts $z=4$, 3, 2, 1.5, and 1. For comparison with other populations of star-forming galaxies, we also show the percentages of non-SMGs (unhatched bars). As expected, both SMGs and non-SMGs are dominant in filamentary environments, accounting for more than 45\% of the population across the redshift range shown here. This is consistent with previous work suggesting that the bulk of galaxies in the universe are in filamentary structures \citep[see e.g.,][]{Cautun.etal.2014, Ganeshaiah_Veena.etal.2019, Pandey.Nandi.2025}. Our results also show that the percentage of SMGs is generally higher than that of non-SMGs in the inner cluster-halo, outer cluster-halo, and inner filament environments, whereas non-SMGs exhibit higher percentages in the outer filament and void/wall environments.

To investigate the evolution of the SMGs population relative to the total galaxy population in the respective cosmic environment, we plot the fraction of SMGs (i.e., $N^{\text{env}}_{\text{SMG}}/N^{\text{env}}_{\text{total}}$) in the middle panel of Fig.~\ref{fig:fractional_evolution} at redshifts $z=4$, 3, 2, 1.5, and 1 using hatched bars. Additionally, for comparison, we show the fraction of non-SMGs (i.e., $N^{\text{env}}_{\text{non-SMG}}/N^{\text{env}}_{\text{total}}$) with unhatched bars.

The inner and outer filament environments exhibit qualitatively similar evolution. In the inner cluster-halo environment, the fraction of SMGs is about $30\%$ at $z=4$, but steadily declines with decreasing redshift, reaching about $3\%$ at $z=1$, corresponding to an order-of-magnitude reduction in the SMG fraction from $z=4$ to $z=1$. In the outer cluster-halo, the SMG fraction stays at about 10\% from $z=4$ to 2, then drops toward lower redshift. By $z=1$, the SMG fraction drops to $\lesssim$2\% in the outer cluster-halo environment.

In both inner and outer filament environments, the SMG fraction shows a modest rise from $z=4$ to 2, followed by a decline to $z=1$, with changes of around $\approx 2$--$3\%$. The maximum SMG fractions in the inner and outer filament regions are $8\%$ and $5\%$ at $z \approx 2$, respectively. The SMG fractions in void/wall environments show a qualitative evolution similar to that in filament environments, a small increase from $z=4$ to 2, followed by a decrease. The maximum SMG fraction in void/wall is even less than $4\%$. The fractional distribution of SMGs in the inner filament, outer filament, and void/wall environments is correlated with the redshift distribution of SMGs (see Figure 3 in \citealt{Kumar.etal.2025} for FLAMINGO). 

In contrast to the results for SMGs, the fraction of the non-SMG population shows a similar evolution in all five environments. It decreases with decreasing redshift for each cosmic environment. The non-SMG population represents a significantly higher fraction of galaxies than the SMGs in any environment. This difference becomes more pronounced as the environmental density decreases.

%Interestingly, if we focus on a population at any given redshift between $z=4$ and 1.5, SMGs and non-SMGs present a distinct density-dependent trend. SMGs are predominantly located in cluster-halo interiors and high-density environments, whereas non-SMGs tend to be more numerous in void/wall regions. The fraction of SMGs in the outer cluster-halo environment is lower than inner filament environment only for $z \leq 1.5$.

%To understand the relative significance of SMGs with respect to non-SMGs in FLAMINGO, we measure the SMG to non-SMG number ratio, $N^{\text{env}}_{\text{SMG}}/N^{\text{env}}_{\text{non-SMG}}$, in a given environment and plot it in the bottom panel of Fig.~\ref{fig:fractional_evolution}. 
To assess the importance of SMGs relative to non-SMGs in a given environment, we compute the number ratio $N^{\text{env}}_{\text{SMG}}/N^{\text{env}}_{\text{non-SMG}}$, shown in the bottom panel of Fig.~\ref{fig:fractional_evolution}.
These findings highlight that SMGs are preferentially located in inner cluster-halos relative to non-SMGs. The ratio of SMGs to non-SMGs in this region drops from about 0.8 at $z=4$ to 0.2 at $z=1$, implying that the inner cluster-halo environment is well represented by SMGs. The ratio of SMGs to non-SMGs in the outer cluster-halo, filament, and void/wall regions is always less than 0.2, suggesting that SMGs sparsely sample these environments compared to non-SMGs across all redshifts shown here. Additionally, in all environments other than the inner cluster-halo, this ratio peaks at $z=2$, consistent with the redshift distribution of SMGs, suggesting that the SMGs best trace large-scale structures at the redshift of their peak activity.

Since SMGs tend to preferentially occupy the inner regions of cluster-halos, our next step is to study how their abundance evolves in massive halos throughout cosmic time. We explore this aspect in the next section.

\subsubsection{Halo occupation evolution of SMGs in massive cluster-halos}
\label{sec:halo_occupation}

\begin{figure}
    \centering
    \includegraphics[width=\columnwidth, height=8in]{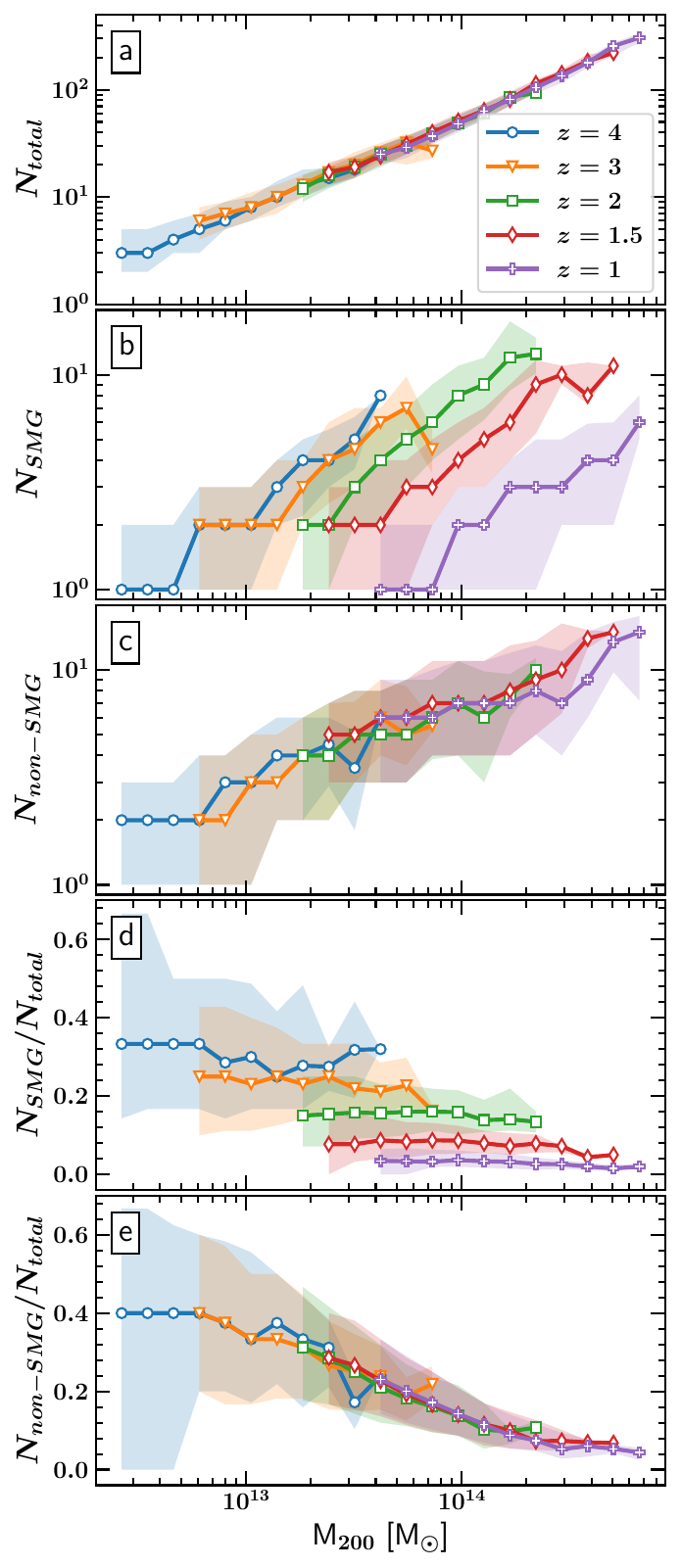}
    \caption{\textit{Panel (a):} The median halo occupation of galaxies within the inner cluster-halo environment at redshifts $z=4$ (blue circles), 3 (orange triangles), 2 (green boxes), 1.5 (red diamonds), and 1 (purple crosses). \textit{Panel (b):} The median halo occupation of SMGs. \textit{Panel (c):} The median halo occupation of non-SMGs. \textit{Panel (d):} The ratio of the halo occupation of SMGs to that of all galaxies. \textit{Panel (e):} The ratio of the halo occupation of non-SMGs to that of all galaxies.
    %The number of SMGs in a cluster-halo increases with mass in proportion to the total number of galaxies. 
    We note that SMGs and non-SMGs are modeled for galaxies with $\rm M_{*} \ge 10^{9}~M_\odot$. In all panels, the shaded regions indicate the 16th to 84th percentile range.}
    \label{fig:halo_occupation}
\end{figure}

%Halo occupation describes the number of galaxies per halo as a function of halo mass \citep[see][and reference therein]{Kravtsov.etal.2004, Zheng.etal.2005}. 
As shown in Section~\ref{sec:fractional_evolution}, SMGs are particularly abundant in dense cluster–halo environments. To examine how the abundance of SMGs varies with halo mass, we compute the halo occupation, i.e., the number of galaxies per halo as a function of halo mass \citep[see][and references therein]{Kravtsov.etal.2004, Zheng.etal.2005}, restricting the analysis to those galaxies in cluster-halo environments. In Fig.~\ref{fig:halo_occupation}, we plot the halo occupation of all galaxies in panel (a), SMGs in panel (b), and non-SMGs in panel (c). The curves with blue circles, orange triangles, green boxes, red diamonds, and purple crosses represent the median fraction of SMGs in halos at redshift $z=4$, 3, 2, 1.5, and 1, respectively. The shaded regions display the 16th to 84th percentile range around the median. The halo occupation of all galaxies and non-SMGs shows similar trends across the redshifts examined for $M_{200}$ in the cluster-halo mass range considered in this study. However, SMGs exhibit clear redshift dependence: high-$z$ halos of a given mass host more SMGs than their low-$z$ counterparts.

%To investigate how the fraction of SMGs in the cluster-halo varies with cluster-halo mass at different redshifts, we calculate the fraction of SMGs in the inner cluster-halo environment (i.e., the ratio of the halo occupation of SMGs to that of all galaxies) and plot it as a function of halo mass in panel (d) of Fig.~\ref{fig:halo_occupation}. 

To assess how the abundance of cluster-halo SMGs scales with overall halo occupation, we compare their halo occupation to that of all cluster-halo galaxies in panel (d) of Fig.~\ref{fig:halo_occupation}. %The curves and shaded regions have the same meaning as in other panels of this figure. 
Our results show that the SMG occupation fraction at a given redshift is nearly constant for M$_{200}$ in the range of cluster-halo masses in this study. The SMG occupation fraction in a halo increases with increasing redshift, with high-redshift halos showing higher SMG occupancy than low-redshift halos. The average median SMG occupation fraction at redshifts $z=4$, 3, 2, 1.5, and 1 are 0.30, 0.23, 0.15, 0.07, and 0.03, respectively.

% https://arxiv.org/abs/1006.0230
% https://arxiv.org/abs/1011.2300

Panel (e) of Fig.~\ref{fig:halo_occupation} shows the same analysis as (d) but for non-SMG galaxies.
%To compare the halo occupation of SMGs with that of non-SMGs in the inner cluster–halo environment, we plot the fraction of non-SMGs (i.e., the ratio of the halo occupation of non-SMGs to that of all galaxies) as a function of halo mass in the bottom panel of Fig.~\ref{fig:halo_occupation}. 
%We use the same colors and symbols for different redshifts as in other panels of this figure. 
In contrast to the constant SMG occupation fraction as a function of halo mass, we find a decreasing non-SMG occupation fraction with increasing halo mass. Additionally, the non-SMG occupation fraction does not evolve from $z=4$ to $z=1$, showing the same median non-SMG occupation fraction at all redshifts for a given halo mass. When going from a halo mass of M$_{200} = 6 \times 10^{12}~\Msun$ to $6 \times 10^{14}~\Msun$, the non-SMG occupation fraction decreases from $40\%$ to $4\%$. This depletion of non-SMGs in the cluster-halo environment is the consequence of a rapidly increasing fraction of star formation suppressed galaxies (SFR < 1 $\MsunYr$) with increasing halo mass. Otherwise, the halo occupation of non-SMGs does not seem to evolve with redshift and qualitatively remains the same even if we relax our SFR threshold for the selection of non-SMGs.

Our results indicate that SMGs preferentially reside in the inner regions of cluster-halos, and their abundance remains roughly proportional to the average number of galaxies per halo at a given redshift. In contrast, non-SMGs are more common in lower-mass halos relative to the average galaxy population and tend to inhabit low-density environments.

\subsection{The properties of SMGs}
\label{sec:properties_SMG}

\begin{figure*}[!ht]
    \centering
    \includegraphics[width=\textwidth]{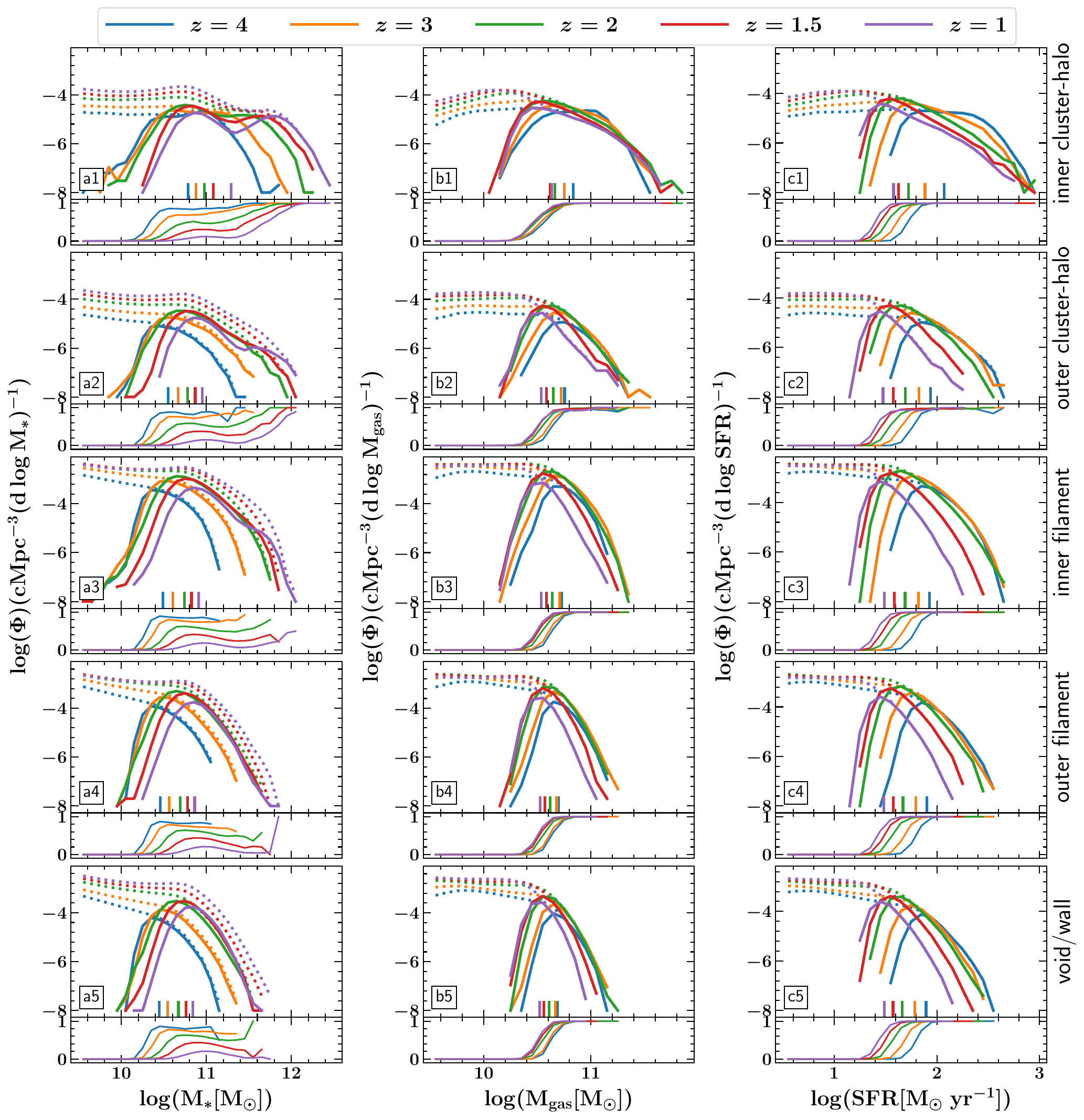}
    \caption{Evolution of the stellar mass function (left panels), star-forming gas mass function (middle panels), and star formation rate function (right panels) in different cosmic environments as indicated on the right-most panel of each row (from top to bottom: inner cluster-halo, outer cluster-halo, inner filament, outer filament, and void/wall). The solid blue, orange, green, red, and purple color curves of each panel represents SMGs in a specific cosmic environment at redshift $z=4$, 3, 2, 1.5, and 1, respectively, whereas the dotted curves of corresponding colors show all galaxies in the same cosmic environment. The vertical marks at the bottom of each panel indicate the mean values of SMGs. The thin panels at the bottom of each main panel shows the relative contribution of SMGs as the ratio of SMGs to total galaxies with $\rm M_{*} \ge 10^{9}~M_\odot$.}
    \label{fig:SMF_GMF_SFRF}
\end{figure*}

To investigate the properties of SMGs in different cosmic environments, we plot their stellar mass function (left panels), star-forming gas mass function (middle panels), and star formation rate function (right panels) for galaxies with $\rm M_{*} \ge 10^{9}~M_\odot$ in Fig.~\ref{fig:SMF_GMF_SFRF}. In each panel, SMGs are shown with solid curves colored blue ($z=4$), orange ($z=3$), green ($z=2$), red ($z=1.5$), and purple ($z=1$), corresponding to the environment indicated in the right-most column of each row. The distribution for the full galaxy sample is shown with dotted lines. To illustrate the relative contribution of SMGs to the total galaxy population, we also show the ratio of SMGs to all galaxies in the thin panels attached at the bottom of each main panel.

The stellar mass function of SMGs typically contributes significantly to the high-mass end of the total stellar mass function across all environments and redshifts (see panels a1–a5). The stellar masses of modeled SMGs agree with their observed median of $\approx 10^{11}\Msun$ \citep{da_Cunha.etal.2015, Michalowski.etal.2017, Miettinen.etal.2017}. 
At fixed redshift, SMGs in dense regions, such as the inner cluster-halo, dominate the massive end of the stellar mass function, with this effect weakening in lower-density environments and becoming less pronounced at $z\simeq 3$–4.
Additionally, their mean stellar mass increases with decreasing redshift, as indicated by the vertical lines at the bottom of each panel. This trend is observable across all environments.

We further find that, in the inner cluster-halo, the stellar mass function of SMGs spans a wider range of stellar masses compared to other environments and exhibits a bimodal distribution. 
Our results show that a significant fraction of satellite galaxies in cluster-halos have submillimeter flux densities exceeding our SMG selection criterion, and are therefore placed in the SMG category. Central SMGs in cluster-halos form the high-mass peak of the stellar mass function, while satellite SMGs contribute to the low-mass peak. This bimodal distribution is also found in outer cluster-halo and inner filament environments, at $z \le 1.5$. In contrast, outer filaments and void/wall environments show no bimodality in the stellar mass function, indicating the absence of satellite SMGs in these regions.

Our analysis of the star-forming gas mass function for SMGs indicates that they possess the largest reservoirs of star-forming gas among galaxies, irrespective of the cosmic web environment and redshift (see panels b1–b5 of Fig.~\ref{fig:halo_occupation}). In large-volume cosmological simulations where the multi-phase interstellar medium is not resolved, star-forming gas is equivalent to cold gas. Therefore, the star-forming gas mass functions indicate that SMGs serve as reservoirs of cold gas ($M_\mathrm{gas} > 10^{10}~\Msun$) at any given epoch. \cite{Tacconi.etal.2008, Bothwell.etal.2013} reported a similar gas mass range in their observational studies. Comparing the mean values (denoted by vertical marks at the bottom) reveals that high-redshift SMGs are generally more abundant in gas than their low-redshift counterparts. In contrast to the stellar mass function, the star-forming gas mass function of SMGs in the inner cluster-halo does not exhibit a bimodal distribution. However, the star-forming gas mass functions in the inner cluster-halo are slightly shallower compared to those in other cosmic environments, except at $z=0$, where the number of highly gas-rich galaxies declines.

Similar to the star-forming gas mass function, the SFR function of SMGs dominates the high-SFR end of the total SFR function across all environments and redshifts (see panels c1–c5 of Fig.~\ref{fig:halo_occupation}). All highly star-forming galaxies are SMGs, regardless of their cosmic environment. At any redshift presented here, SMGs consistently exhibit an SFR $\gtrsim 15~\MsunYr$, with the smallest mean SFR exceeding $30~\MsunYr$. High-redshift SMGs are generally more star-forming than their low-redshift counterparts, as evidenced by the mean values indicated at the bottom of each panel. 

\begin{figure*}[!ht]
    \centering
    \includegraphics[width=\textwidth]{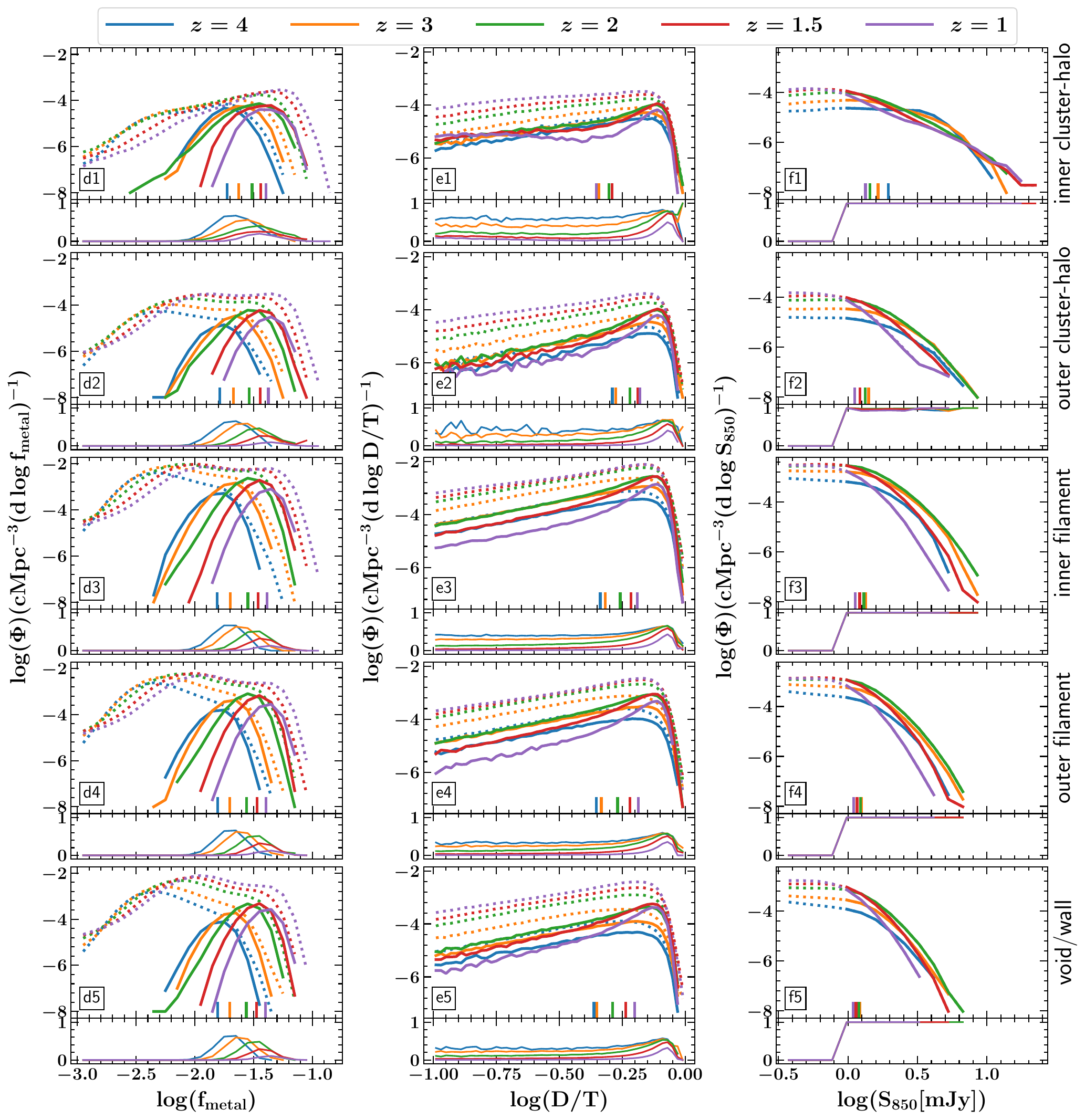}
    \caption{Evolution of the star-forming gas metallicity function (left panels), disk-to-total mass ratio function (middle panels), and submillimeter flux density function (right panels) in different cosmic environments as indicated on the right-most panel of each row. The solid blue, orange, green, red, and purple color curves of each panel represents SMGs in a specific cosmic environment at redshift $z=4$, 3, 2, 1.5, and 1, respectively, whereas the dotted curves of corresponding colors show all galaxies in the same cosmic environment. The vertical marks at the bottom of each panel indicate the mean values of SMGs. }
    \label{fig:SFMetF_DTF_FDF}
\end{figure*}

To further investigate the nature of SMGs in different cosmic environments, we plot the star-forming gas metallicity (${\rm f}_{\rm metal}$) function of galaxies (left panels), the disk-to-total stellar mass ratio function (middle panels), and the $S_{850}$ flux density function (right panels) in Fig.~\ref{fig:SFMetF_DTF_FDF}. In each panel, the solid blue, orange, green, red, and purple curves represent SMGs at redshifts $z=4$, 3, 2, 1.5, and 1, respectively. The corresponding dotted curves show the full galaxy population in each environment.
Vertical lines at the bottom of each panel indicate the mean values for SMGs.

The star-forming gas metallicity functions across redshifts and environments indicate that SMGs tend to be metal-rich (see panels d1-d5 in Fig.~\ref{fig:SFMetF_DTF_FDF}). This is expected given their dusty nature, as the submillimeter flux in our modeling depends on the dust mass, which is directly derived from the metal mass, assuming a fixed dust-to-metal ratio. However, not all highly metal-rich galaxies are SMGs. A considerable number of metal-rich galaxies have exhausted their gas reservoirs and exhibit low star formation rates (SFR $<10~\MsunYr$). In contrast, many SMGs host dust that is diluted by large gas reservoirs, reducing their apparent metal richness. This analysis shows that while SMGs are dusty, they are not necessarily metal-rich.
Furthermore, the contribution of SMGs to the metal-rich end of the star-forming gas metallicity function increases with redshift across all environments, suggesting that metal-rich galaxies are more likely to be SMGs at higher redshift. Due to the chemical enrichment of the universe, the mean metal fraction in SMGs increases with decreasing redshift, as indicated by the vertical lines at the bottom of each panel \citep[also see observational study by][and reference therein]{Bethermin.etal.2015}. This increasing metal mass offsets the effect of SFR depletion on the submillimeter flux in galaxies at lower redshifts. The star-forming gas metal fraction functions of SMGs remain approximately constant across different environments at any given epoch.

We further investigate the galaxy morphology of SMGs according to their environment. For this, we use the fraction of co-rotating stellar mass as a proxy for the disk-to-total stellar mass ratio ($D/T$)\footnote{The disk-to-total ratio is defined as $D/T = 1 - 2 (M_{\rm count-rot} / M_{\rm total})$, where $M_{\rm count-rot}$ is the mass of counter–rotating particles with respect to total angular momentum and $M_{\rm total}$ is the total stellar mass within 50~pkpc spherical aperture.} Given the scale of the simulation and its mass resolution, a detailed photometric or kinematic decomposition is unfeasible. Although the co-rotating stellar mass fraction is not a highly reliable morphological indicator for galaxies modeled with a small number of particles, it serves as a useful parameter for statistical analysis. Across redshifts and cosmic environments, the $D/T$ functions of SMGs and all galaxies show a peak near $D/T \sim 0.4-0.5$ (see panels e1-e5 of Fig.~\ref{fig:SFMetF_DTF_FDF}), where galaxies exhibit disk-like kinematics. 

The mean SMG values are also shown at the bottom of each panel.
In any given environment, the mean value increases with decreasing redshift, suggesting that SMGs are more disky as the universe evolves. At $z = 3$ and $4$, $D/T$ is very close to 0.5, while $D/T$ is greater than 0.6 at $z = 2$ to 1. We note that the $D/T$ ratio in the outer cluster-halo environment is greater than 0.6 for all the redshifts shown here. Furthermore, we find that for SMGs, the $D/T$ distribution is less uniform and more sharply peaked than that of the overall galaxy population.

From an observational perspective, several studies have shown that SMGs exhibit disk-like or rotationally supported morphologies. Recent JWST imaging reveals that many SMGs are extended systems with irregular or clumpy disk structures rather than compact spheroids or clear major mergers \citep{Gillman2023,Chan2025,Umehata2025}. In particular, \citet{Umehata2025} investigate SMGs in protocluster environments, providing observational evidence that massive dusty star-forming galaxies in overdense regions can display extended and structured morphologies.
Within this context, the predominance of SMGs with $D/T \geq 0.5$ found in our simulations is broadly consistent with observations. The more sharply peaked and less uniform $D/T$ distribution for SMGs relative to the overall galaxy population may indicate that SMGs constitute a more homogeneous subset of massive, gas-rich star-forming galaxies. 

Although no current observational study probes the morphology of SMGs as a function of their location in the cosmic web, our results provide predictions for future observational work combining high-resolution imaging, kinematics and large-scale environment reconstruction.  

To probe environmental effects on submillimeter emission, we present S$_{850}$ flux density distributions across distinct cosmic environments. (see panels f1–f5 of Fig.~\ref{fig:SFMetF_DTF_FDF}). We find that the adopted SMG flux limit is typically set above the knee of the flux density distribution. \cite{Kumar.etal.2025} found that SMGs increase in number from high redshift to $z=2.5$ in FLAMINGO as a whole population. We observe a similar growth in number from high redshift down to $z=2$ in all environments except the inner cluster-halo. The inner cluster-halo SMGs show nearly identical flux density functions from $z=3$ to 1, with the $z=4$ flux density function showing a slight deficit on both sides of its knee. This suggests that SMGs are brighter in the cluster-halo environment earlier than in any other environment and maintain high fluxes in the redshift range shown here. On the other hand, other environments show a sharp decrease in bright SMGs after $z=2$, consistent with the overall flux density functions shown in \cite{Kumar.etal.2025}. This suggests that the main contributors to the evolution of the complete flux density functions beyond the knee are the SMGs in the outer cluster-halo, filament, and void/wall regions.

Notably, the inner cluster-halo hosts the brightest galaxies, followed by the outer cluster-halo, then the inner filament, the outer filament, and finally the void/wall regions. In FLAMINGO, only the inner cluster-halo hosts SMGs brighter than 10 mJy, suggesting that brighter SMGs are typically associated with high-density environments. The flux density functions in the filament and void/wall environments show quantitatively lower values than those in the cluster-halo environments near S$_{850}=1$ mJy, indicating that a large population of faint SMGs resides in low-density regions.

\subsection{The contribution of SMGs to cosmic star formation}\label{sec:contribution_to_SFRD}

\begin{figure}
    \centering
    \includegraphics[width=\columnwidth]{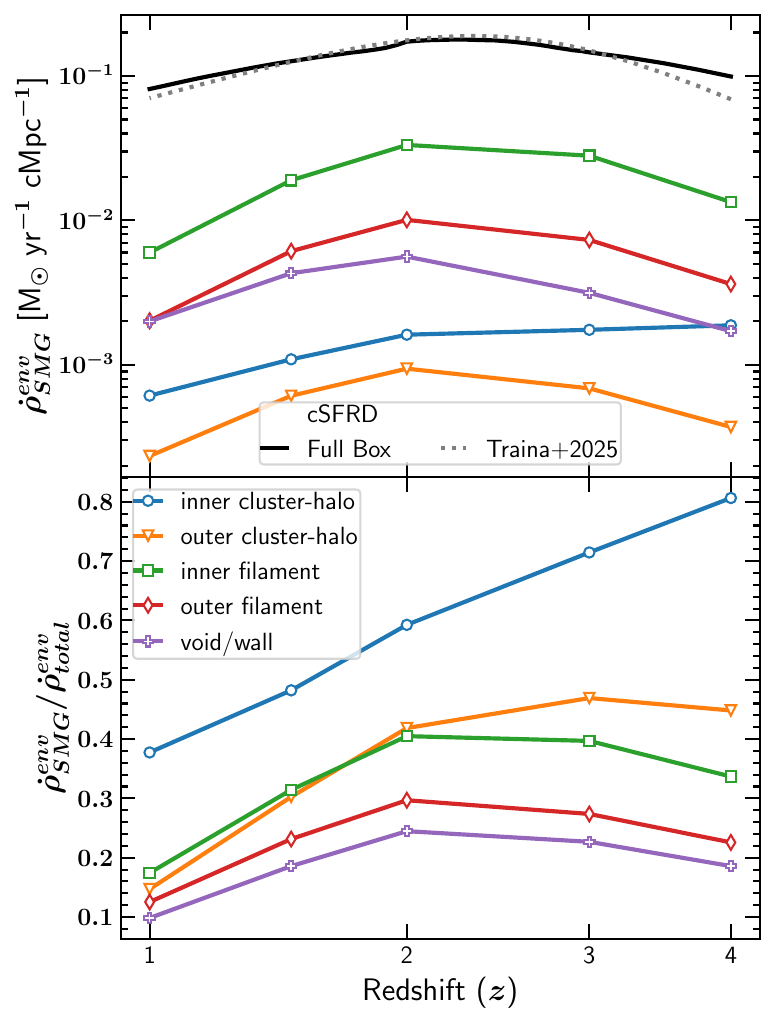}
    \caption{\textit{Top:} The contribution of SMGs to the cosmic SFRD from the inner cluster-halo (blue circles), outer cluster-halo (orange down triangle), inner filament (green boxes), outer filament (red diamonds), and void/wall (purple crosses) environments as a function of redshift. We also show the full box FLAMINGO SFRD (solid black) and \cite{Traina.etal.2025} fit to observation (dotted gray) for comparison. \textit{Bottom:} The contribution of SMGs to the SFRD in five environments as a function of redshift. The ratio represents SFRD of SMGs to that of all galaxies in the respective environment for galaxies in this study (M$_{*} \ge 10^{9}~\Msun$). Inner filament environment dominates the cosmic SFRD contributed by SMGs, whereas the environment in which SMGs contribute the most to the environmental SFRD is the inner cluster-halo.
    }
    \label{fig:contribution_to_SFRD}
\end{figure}

SMGs contribute significantly to the cosmic star formation rate density. At $z=2$--3, about $30\%$ of the cosmic star formation comes from SMGs with S$_{850} >$ 1 mJy \citep{Dudzeviciute.etal.2020, Kumar.etal.2025}. Here, we analyze their contribution within the distinct cosmic environments established in this work. In the top panel of Fig.~\ref{fig:contribution_to_SFRD}, we show the contribution of SMGs to the cosmic SFRD across different environments. The blue, orange, green, red, and purple curves represent the inner cluster-halo, outer cluster-halo, inner filament, outer filament, and void/wall environments, respectively. For reference, we also plot the full-box SFRD from FLAMINGO (solid black curve) and a recent fit to observational SFRD (dotted gray curve) from \cite{Traina.etal.2025}. We note that the full-box SFRD from FLAMINGO includes all galaxies, including those with stellar masses below the range considered in this study (i.e., M$_{*} < 10^{9}~\Msun$). Although FLAMINGO is not explicitly calibrated for the cosmic SFRD, its predicted SFRD agrees well with observational estimates across the redshift range examined here. Consistent with the distribution of SMGs across environments shown previously in Fig.~\ref{fig:fractional_evolution}, the inner filament environment dominates the SMG contribution to the cosmic SFRD, followed by the outer filament, void/wall, inner cluster-halo, and outer cluster-halo environments over the redshift range considered.
At the peak of SMG activity, around $z \approx 2$, we find that only a small fraction of the total contribution comes from the cluster-halo environments: the inner and outer regions together account for just $\approx 1.5\%$. The more diffuse void and wall environments contribute a slightly larger share, at $\approx 3\%$. Moving to denser large-scale structures, the outer filament regions provide $\approx 6\%$ of the total, while the inner filaments dominate with a substantial $\approx 19\%$ contribution. This enhancement in filamentary environments is not surprising, as a large fraction of SMGs are preferentially located within filaments (see Section~\ref{sec:fractional_evolution}).

Further, we investigate the relative contribution of SMGs to the SFRD from galaxies considered in the study (i.e., M$_{*} \ge 10^{9}~\Msun$), thereby extending our analysis from the global SFRD to the specific role of SMGs across environments. In the bottom panel of Fig.~\ref{fig:contribution_to_SFRD}, we show the fractional SFRD evolution of SMGs in different environments relative to the total SFRD in the respective environment. The contribution of SMGs in the inner cluster-halo is the highest at all redshifts shown here. Their contribution in the inner cluster-halo varies from $40\%$ at $z=1$ to $80\%$ at $z=4$. Thus, the contribution of SMGs to the environmental SFRD grows systematically with increasing density. SMGs in the densest regions, such as the inner cluster-halo, dominate the SFRD contribution, whereas those in lower-density environments, such as void/wall, contribute the least. The only exceptions are the outer cluster-halo and inner filament at $z \leq 2$, where they show approximately the same contribution from SMGs.

\section{Conclusions}
\label{sec:conclusions}
In this work, we utilize the large-volume cosmological hydrodynamical simulation FLAMINGO, which simultaneously reproduces the redshift distribution and number counts of SMGs without invoking a top-heavy initial mass function. Considering galaxies ($M_{*} \geq 10^{9}~\Msun$) as tracers of large-scale structure, we use the DisPerSE cosmic web tracer to identify the cosmic web structure at $z=4$, 3, 2, 1.5, and 1. We opt for a stellar-mass–selected galaxy sample to more closely match observational cosmic web reconstructions.
We define five cosmic environments (see Fig.~\ref{fig:env_definition}): inner cluster-halo, outer cluster-halo, inner filament, outer filament, and void/wall, and associate SMGs with these regions to investigate their evolution across different cosmic environments. 

To mitigate the effects of cosmic evolution, we measure the comoving number density of halos with $M_{200} \geq 10^{14}~\Msun$ at $z = 0$ (which is $9.736 \times 10^{-6}$ cMpc$^{-3}$ in the FLAMINGO simulation). We use this value to define the minimum cluster-halo mass at all redshifts, assuming that the comoving number density of cluster-halos remains constant. 
Although the most massive halos at any given redshift do not necessarily evolve into $z = 0$ cluster-halos, this method instead follows regions of similar overdensity over cosmic time, thereby pinpointing likely progenitors of today’s massive clusters without enforcing a redshift-independent halo mass cut. For filaments, we compute the stacked galaxy number density profile perpendicular to the filament spine and find negligible evolution in physical space across redshifts. Consequently, we adopt a fixed physical radius of 1 pMpc for the inner filament and $1$--$2$ pMpc for the outer filament environment. We use SCUBA-2 S${850}$ band flux densities from \citet{Kumar.etal.2025} and define galaxies with S${850} > 1$~mJy as SMGs. For comparison, we define the non-SMG population (serving as a proxy for other star-forming galaxy populations, such as LAEs and LBGs) as galaxies with S$_{850} < 0.5$ mJy and SFR $> 2~\MsunYr$. The following summarizes our analysis and key findings.

\begin{enumerate}
\item At $z = 4$, about $30\%$ of galaxies with M$_{*} \ge 10^{9}~\Msun$ in the inner cluster-halo environment are SMGs, but this fraction declines sharply to $\sim 3\%$ by $z = 1$ (see Fig.~\ref{fig:fractional_evolution}). In the outer cluster-halo, the SMGs remains around $10\%$ from $z = 4$ to 2, then falls below $2\%$ by $z = 1$. In the inner filament, outer filament, and void/wall environments, the SMG fraction rises slightly from $z = 4$ to 2, then declines toward $z = 1$. In contrast, the non-SMG population exhibits a similar evolutionary trend across all environments, consistently decreasing with redshift. For $z > 1.5$, SMGs show a clear environmental gradient, with their fraction decreasing from inner cluster-halo to outer cluster-halo, inner filament, outer filament, and finally void/wall regions.

\item The number of SMGs in a cluster-halo increases with mass in proportion to the total number of galaxies of M$_{*} \ge 10^{9}~\Msun$ (see Fig.~\ref{fig:halo_occupation}). Thus, at a given redshift, the ratio of the halo occupation of SMGs to that of all galaxies in cluster-halos is nearly independent of cluster-halo mass but increases with redshift, with median values of 0.30, 0.23, 0.15, 0.07, and 0.03 at $z = 4$, 3, 2, 1.5, and 1, respectively. In contrast, the ratio of the halo occupation of non-SMGs  to that of all galaxies show a decreasing trend with increasing halo mass and no significant redshift evolution. This depletion of non-SMGs in massive halos is driven by a rising population of weakly star-forming galaxies (SFR $< 1~\MsunYr$).

\item For a given redshift and environment, SMGs represent massive, gas-rich, star-forming, metal-rich, and disky populations. Denser environments show more massive, gas-rich, star-forming and luminous SMGs. At $z=4$--1, SMGs show mean stellar mass $\log (M_{*}/\Msun) \geq 10.4$, mean gas mass $\log (M_{g}/\Msun) > 10.5$, mean SFR $\gtrsim 30~\MsunYr$, and mean star-forming gas metal fraction f$_{\rm metal} \geq 0.016$ (see Fig.~\ref{fig:SMF_GMF_SFRF} and ~\ref{fig:SFMetF_DTF_FDF}).

\item At a given redshift, SMG contributions to the high-mass end of the stellar mass function decline from dense inner cluster-halos to voids/walls. This contribution also drops with decreasing redshift across all environments due to suppressed star formation in massive galaxies. In inner cluster-halos, SMGs span a broader stellar mass range and show a bimodal distribution: central SMGs dominate the high-mass peak, while satellite SMGs contribute to the low-mass peak.

\item Across redshifts and environments, SMGs dominate the metal-rich end of the star-forming gas metal fraction functions, consistent with their dusty nature. However, not all metal-rich galaxies are SMGs--many have depleted gas reservoirs and low star formation rates (SFR $< 10~\MsunYr$).

\item The brightest SMGs at all epochs reside in the inner cluster-halo, followed by outer cluster-halo, inner filament, outer filament, and void/wall environments. In FLAMINGO, only inner cluster-halos host SMGs brighter than 10 mJy, indicating a strong link between SMG brightness and high-density environments.

\item Across all redshifts, the contribution of SMGs to the star formation occurring in a particular type of region is largest for inner cluster regions (see Fig.~\ref{fig:contribution_to_SFRD}). In other environments, their contribution remains below $50\%$. The SFRD fraction of SMGs in any environment correlates with environmental density, decreasing from inner cluster-halos to outer cluster-halos, inner filaments, outer filaments, and voids/walls.

\end{enumerate}

Building on these findings, future work would benefit exploring the physical mechanisms driving the environmental dependence of SMG evolution, particularly the role of cluster-halo-specific processes such as mergers and AGN feedback in shaping their star formation histories. High-resolution simulations and multi-wavelength observations, especially in the submillimeter and infrared, will be crucial to disentangle the interplay between gas accretion, metallicity enrichment, and morphological transformation across cosmic time.

\begin{acknowledgements}
We thank D. Galárraga-Espinosa for her suggestions. AK and MCA acknowledge support from ALMA fund with code 31220021 and from ANID BASAL project FB210003. ADMD acknowledge support from the Universidad Técnica Federico Santa María through the Proyecto Interno Regular \texttt{PI\_LIR\_25\_04}. LG also gratefully acknowledges financial support from ANID - MILENIO - NCN2024\_112, ANID BASAL project FB210003, FONDECYT regular project number 1230591. HSH acknowledges the support of the National Research Foundation of Korea (NRF) grant funded by the Korea government (MSIT), NRF-2021R1A2C1094577, and Hyunsong Educational \& Cultural Foundation. J.L. is supported by the National Research Foundation (NRF) of Korea grant funded by the Korea government (MSIT, RS-2021-NR061998). SL acknowledges support from the National Research Foundation of Korea(NRF) grant  (RS-2025-00573214) funded by the Korea government(MSIT).
%Tools
We make use of Matplotlib \citep{matplotlib2007}, Numpy \citep{numpy2020}, Pandas \citep{pandas2010, pandas2020}, Scipy \citep{Scipy2020}, WebPlotDigitizer in this work. 
%RAGNAR
This work used the RAGNAR computing facility available at Universidad Andrés Bello.
%COSMA
This work used the DiRAC@Durham facility managed by the Institute for Computational Cosmology on behalf of the STFC DiRAC HPC Facility (\url{www.dirac.ac.uk}). The equipment was funded by BEIS capital funding via STFC capital grants ST/P002293/1, ST/R002371/1 and ST/S002502/1, Durham University and STFC operations grant 
ST/R000832/1. DiRAC is part of the National e-Infrastructure.
\end{acknowledgements}
\bibliographystyle{aa}
\bibliography{references.bib}
\end{document}